\documentclass[
aps,nofootinbib,
showpacs,showkeys,preprint
tightenlines,preprintnumbers,] {revtex4}

\usepackage{epsf,epsfig,subfigure,graphicx,amsmath,amssymb 
}
\usepackage{color}
\newcommand{\ie}{{\it i.e.~}}

\newcommand{\Qem}{Q_{\rm em}}

\def\sw0{{$\sin^2\theta_W^0$}}

\def\fGUT{{\rm family-GUT}}

\newcommand{\Z}{{\bf Z}}

\def\E6{{\rm E_6}}

\def\EE8{{\rm E_8\times E_8'}}

\def\anti{anti-SU$(5)$}

\def\six{{\bf 6}}

\def\nine{{\bf 9}}
\def\nineb{\overline{\bf 9}\,}

\def\tsixb{\overline{\bf 36}\,}
\def\tsix{{\bf 36}}

\def\one{{\bf 1}}

\def\two{\bf 2}
\def\five{{\bf 5}}
\def\ten{{\bf 10}}
\def\tenb{{\overline{\bf 10}}} 

\def\fiveb{{\overline{\bf 5}}}
\def\threeb{{\bf\overline{3}}}
\def\three{{\bf 3}}

\begin{document}

\draft

\title{\Large\bf R-parity  from string compactification}

\author{ Jihn E.  Kim}
\address
{Department of Physics, Kyung Hee University, 26 Gyungheedaero, Dongdaemun-Gu, Seoul 02447, Republic of Korea}

\begin{abstract} 
The strategy for assigning $\Z_{4R}$ parity in the string compactification is presented. For the visible sector, an anti-SU(5) (flipped-SU(5)) grand unification (GUT) model with three families is used to reduce the number of representations compared to the number in the minimal supersymmetric standard models (MSSMs). The SO(32) heterotic string  is used to allow a large nonabelian gauge group SU($N$), $N\ge 9$, for the hidden sector such that the number of extra U(1) factors is small. A discrete subgroup of the gauge U(1)'s is defined as the  $\Z_{4R}$ parity.  Spontaneous symmetry breaking of anti-SU(5) GUT is achieved by the vacuum expectation values of two index antisymmetric tensor Higgs fields $\ten_{+1}$ and $\tenb_{-1}$ that led to our word `anti-SU(5)'.  In the illustrated example, the multiplicity 3 in  one twisted sector allows the permutation symmetry $S_3$ that leads us to select the third family members and one MSSM pair of the Higgs quintets.  
 
\end{abstract}
\maketitle


\section{Introduction}\label{sec:Introduction}
\noindent

Grand unified theories (GUTs) attracted a great deal of attention ethetically because they provided unification of gauge couplings and charge quantization \cite{GQW74,PS73,GG74}. But there seems to be a fundamental reason leading to GUTs even at the standard model (SM) level.   With the electromagnetic and charged currents (CCs), the leptons need representations which are a doublet or bigger. A left-handed (L-handed) lepton doublet  $(\nu_e,e)$ alone is not free of gauge anomalies because the observed electromagnetic charges are not $\pm \frac12$. The anomalies from the fractional electromagnetic charges of the $u$ and $d$ quarks are needed to make the total anomaly to vanish \cite{Bouchiat72,Jackiw72}. In view of this necessity for jointly using both leptons and quarks to cancel gauge anomalies even in the SM, we can view that unification of leptons and quarks is fundamentally needed in addition to the esthetic view.

In the SM, the largest number of parameters arises from the Yukawa couplings which form the bases of the family structure. Repetition of fermion families in 4-dimensional (4D) field theory or  family-unified GUT (\fGUT) was formulated by Georgi \cite{Georgi79}, requiring un-repeated chiral representations while not allowing gauge anomalies. Some interesting \fGUT\,models are the spinor representation of SO(14) \cite{Kim80PRL,Kim81PRD} and ${\bf 84}\oplus 9\cdot\overline{\bf 9}$ of SU(9) \cite{Frampton80}.\footnote{For more attempts of \fGUT s, see references in \cite{FramptonKim20}.} While Refs. \cite{Kim80PRL,Kim81PRD,Frampton80}  do not provide interesting non-vanishing flavor quantum number, the SU(11) model \cite{Georgi79} allows a possibility for non-vanishing flavor quantum number such as U(1)$_{\mu-\tau}$ or U(1)$_{B-L}$ \cite{FramptonKim20}.

\bigskip
\noindent
On the other hand, the standard-like models from string have been the main focus of phenomenological activities for the ultraviolet completion of the SM toward the minimal supersymmetric standard model (MSSM) in the last several decades \cite{IKNQ, Munoz88, Lykken96, PokorskiW99, Cleaver99, Cleaver01, CleaverNPB, Donagi02,  Raby05, Donagi05, He05,  Donagi06, He06, Blumenhagen06, Cvetic06, Blumenhagen07, KimJH07, Faraggi07, Cleaver07, Munoz07,Lebedev08, Nilles08,Nibbelink07, Vaudrevange11}. These models use the chiral specrum from the level--1 construction which leads to unification of gauge couplings \cite{Ginsparg87}.
So, the standard-like models from string compactification achieved the goal of gauge coupling unification and GUT theories from string  have not attracted much attention. Nevertheless, GUTs from strings \cite{Ellis89,KimKyae07} have been discussed sporadically  for anti-SU(5) \cite{DKN84} (or flipped SU(5) \cite{Barr82}), dynamical symmetry breaking \cite{Huh09,KimKyae19},  
and family unification \cite{Kim19Rp}. In fact, \fGUT s  are much easier in discussing the family problem, in particular on the origin of the mixing between quarks/leptons because the number of representations in \fGUT s is generally much smaller than in their (standard-like model) subgroups.

\bigskip
\noindent
In this paper we study the R-parity assignment for a \fGUT~from string compactification. So far, most  string compactification models used the $\EE8$ heterotic string in which a GUT with rank greater than 8 is impossible.  The group SU(11) has rank 10   which cannot arise from compactification of $\EE8$. Therefore, firstly we fomulate the orbifold compactification \cite{DHVW1,DHVW2} of SO(32) heterotic string \cite{GHMR1} whose rank is 16.   From the compactification of the SO(32) heterotic string, however, we cannot derive the SU(11) model and  the largest possible non-abelian gauge group we obtain is  SU(9) \cite{KimIJMPA20}.

\bigskip
\noindent
To discuss the R-parity from string compactification, we should consider a specific model. Among compactification schemes, we adopt the orbifold method. Among 13 possibilities listed in Ref. \cite{DHVW1,DHVW2}, we employ $\Z_{12-I}$ orbifold because it has the simplest twisted sectors. Twisted sectors are distinguished by Wilson lines \cite{INQ87}. The Wilson line in $\Z_{12-I}$ distinguishes three fixed points at a  twisted sector. Therefore, it suffices to consider only three cases at a twisted sector. In all the other orbifolds of Ref. \cite{DHVW1}, consideration of various possibilities of Wilson lines and the accompanying consistency conditions are much more involved. 
 
\bigskip    
 \noindent 
 In Sec. \ref{section:Orb}, we recapitulate the orbifold methods used in this paper for an easy reference to  Sec. \ref{sec:SU9}.  
 In Sec. \ref{sec:SU9}, we construct a specific specific supersymmetric model possessing nonabelian groups SU(9) and SU(5).  In Sec. \ref{sec:Rparity}, we assign $\Z_{4R}$ quantum numbers to the massless fields obtained in Sec.  \ref{sec:SU9}.   Section \ref{Conclusion} is a conclusion.

\section{Orbifold compactification}\label{section:Orb}

\subsection{Tensor representations of SU($N$) and SO($2N$) spinors}\label{sec:SU16}

\bigskip 
\noindent 
All anti-symmetric representations $\Phi^A, \Phi^{AB}\equiv \Phi^{[AB]}, \cdots  \Phi^{[AB\cdots]}$, etc. of SU($N$) contain only $\three$ and $\threeb$ under its subgroup SU(3). Therefore, using only anti-symmetric representations for matter in an SU($N$) GUT guarantees no high dimensional colored matter particles such as $\six,\bar{\six}$, etc. The SU($N$) representations are either of vector-type or of spinor type. Let us define the vector representation of SU($N$) as 
\begin{eqnarray}
{\rm SU(9):~}\nine=(\underline{-1\,0^8})  ;~ 
\tsix=(\underline{-1\,-1~~0^7})\label{eq:Vector9}\\
{\rm SU(5):~}\five=(\underline{+1\,0^4})  ;~ 
\ten=(\underline{+1\,+1~~0^3})\label{eq:Vector5}
\end{eqnarray}
where we defined differently for SU(9) and SU(5).  It is a matter of convention to call what are {\bf 9} and {\bf 5}.  Here, the first (second) ones are called one-(two-)index tensors. The second ones carry two non-zero numbers. For spinor types, we also use this convection, but spinors fill all the 9 (for SU(9)) and 5 (for SU(5)) slots such as $(\underline{++--\cdots})$. One index spinor is defined to have one sign and the rest the opposite sign. In string compactification, there appear only up to two index tensors. Define $\tsixb$ of the spinor type in SU(9) such that $\tsixb\cdot\nine\cdot\nine$ is allowed if two times the sum of entries is 0 modulo 9.\footnote{The spinor $(+++++++++)$ for $+\equiv \frac12$ gives two times the sum as 9.} Therefore, we have    $\tsixb = (\underline{++-------})$.
Similarly, we have the spinor $\tenb$ of SU(5) as
$\tenb=(\underline{+++--}) $, satisfying +1+2+2=5 from $\tenb\cdot\five\cdot\five$. Thus, one and two index spinors are
\begin{eqnarray}
&{\rm SU(9):~}\nineb=(\underline{+--------})  ;~ 
\tsix=(\underline{--+++++++}).\label{eq:Spinor9}\\
&{\rm SU(5):~}\five=(\underline{+----})  ;~ 
\ten=(\underline{++---}) \label{eq:Spinor5}
\end{eqnarray}
One $\tsix$ of (\ref{eq:Vector9}) and two $\nineb$'s of (\ref{eq:Spinor9}) give $ -4-7-7=-18$ from $\tsix\cdot\nineb\cdot\nineb$.

\bigskip 
\noindent 
Counting the dimension of SO(32) spinor, we note that both even and odd numbers of + signs should be included. This is in contrast to the E$_8$ spinor where  only even numbers of + signs are considered.

\subsection{Two dimensional orbifolds}\label{subsec:OrbTwo}
\noindent
In compactifying 10D string theory to 4D effective field theory, the small internal 6D is a compactified three two-tori. So, the basic is a two dimensional torus. A two dimensional orbifold is this two dimensional torus modded by discrete groups for which we adopt the discrete group $\Z_{12-I}$. The shift vector we use here is similar to that of  \cite{KimIJMPA20} but not the same, and the spectrum we obtain is a bit simpler. Since the strategy of compactifying SO(32) heterotic string is already presented there, we list key formulae used in the next chapter in Appendix \ref{app:Fixed}.

\bigskip    
 \noindent 
Most interesting spectra in the present paper are arising in the twisted sectors. Geometrically, twisted sectors correspond to fixed points. The moding vector of $\Z_{12-I}$ is $\phi_s$  \cite{DHVW1,DHVW2},
\begin{equation}
\phi_s=\frac{1}{12}(5,4,1).\label{eq:Z12I}
\end{equation}
The multiplicities in the fixed points are three because $\phi_s$ contains $\frac13$ in the second torus. This multiplicity 3 can be distinguished by Wilson lines $a_3=a_4$ (3, 4 denoting two directions in the second torus) \cite{INQ87}. Entries of $a_3$ are some integer multiples of $\frac13$. But $3a_3$ contains only integer entries, that leads to some conditions at twisted sectors $T_3, T_6$ and $T_9$.

 \bigskip
 \noindent
 There is one point to be noted for $\Z_{12}$. If $N=$even, the $k=1,\cdots,\frac{N}{2}-1$ sectors provide the opposite chiralities in the  $k=N-1,\cdots,\frac{N}{2}+1$ sectors, 
 \begin{equation}
 T_k\leftrightarrow T_{N-k}.\nonumber
\end{equation}
Then, the corresponding phases of Eq. (\ref{eq:Phase}) compare as
 \begin{equation}
 e^{2\pi i \Theta_k}\leftrightarrow  e^{2\pi i \Theta_{N-k}}\label{eq:PhaseDiff}
\end{equation}
whose difference is $e^{2\pi i(N-2k)/12}$. Thus, if $2k=N$ then $T_{N-k}$ do not provide the charge conjugated fields of $T_k$, but they are identical. For $T_3$ or $T_9$, therefore, we must provide the additional charge conjugated fields with an extra phase $e^{2\pi i(10/12)}=e^{2\pi i(-2/12)}$, the difference of $\hat{s}\cdot \phi_s$ for $\hat{s}=(+++)$ of R and $(-\,-\,-)$ of L.  We choose $T_9$ for this phase. In $T_6$, however,  we do not need this since the charged conjugated fields also appear there.    
 
\subsection{Multiplicities in the twisted sectors} 

\noindent 
 In the compactification of  the  SO(32) heterotic string, spinors in $U$ are not appearing because it is not possible to have $P^2=2$ from sixteen $\pm\frac12$'s.    Only vector types are possible in $U$.  The $T_k$ twisted sector has three possibilities
\begin{eqnarray}
T_k^{0,+,-}: ~kV_a=\left\{\begin{array}{l}kV\equiv kV_0\\[0.2em]
k(V+a_3)\equiv kV_+\\[0.2em]
k(V-a_3)\equiv kV_-.\end{array}\right.
\end{eqnarray} 

\bigskip
\noindent
We select only the even lattices shifted from the untwisted lattices, therefore, we consider even numbers for the sum of absolute value of each element of $P$.  They should be even numbers if the absolute values  are added.   In Appendix \ref{app:Fixed}, $\Theta_k$ in Eq. (\ref{eq:Phase}) is defined for the twisted sector $T_k$. Since different  $P$'s in the same twisted sector may lead to  different gauge group representations,   there is a need to distinguish them. So, we may use
\begin{eqnarray}
&\Theta_{\rm Group}  =-\tilde{s}\cdot\phi_s -k\,p_{\rm vec}^{ k\,\rm th}\cdot \phi_{s} + k\,P\cdot V_0+\frac{k}{2}(\phi_s^2-V_0^2)+\Delta_k^N-  \delta_k^N ,\label{eq:PhaseA}
\end{eqnarray}
where 
\begin{eqnarray}
\delta_k^N=2\delta_k .\label{eq:deltakN}
\end{eqnarray}

\section{The model}
\label{sec:SU9}
 
\bigskip
\noindent 
The left-hand side (LHS) sector of the heterotic string is the gauge sector. The shift vector $V_0$ and Wilson line $a_3$ are restricted to satisfy the $\Z_{12-I}$ orbifold conditions, 
\begin{eqnarray} 
12 (V_0^2-\phi_s^2)= 0~\textrm{mod even integer},~
 12(V_0\cdot a_3)=0~\textrm{mod even integer},~
12|a_3|^2 =0~\textrm{mod even integer}. \label{eq:Conditions}
\end{eqnarray}
Here, $a_3\,(=a_4)$ is chosen to allow and/or forbid some spectra,
and is composed of fractional numbers with the integer multiples of $\frac13$ because the second torus has the $\Z_3$ symmetry.  
The model is
\begin{eqnarray}
 \begin{array}{l}  V_0=\left(\frac{1}{12},\,\frac{1}{ 12},\,\frac{1}{12},\,\frac{1}{12},\,\frac{1}{12},\,\frac{1}{12},\,\frac{1}{ 12},\,\frac{1}{12},\,\frac{1}{12};\, \frac{3}{12},\,  \frac{6}{12}\,;\, \frac{6}{12},\,\frac{6}{12},\,\frac{6}{12},\,\frac{6}{12},\,\frac{6}{12} \right) ,~V^2_0=\frac{234}{144}\to \frac{-54}{144},    \\[0.4em]
 V_+=\left( \frac{+1}{12},\, \frac{+1}{12},\, \frac{+1}{12}, \, \frac{+1}{12},\, \frac{+1}{12}, \,\frac{+1}{12},\, \frac{+1}{12},\, \frac{+1}{12}, \,\frac{+1}{12};\, \frac{-5}{12},\, \frac{+6}{12}\,;\, \frac{+2}{12},\, \frac{+2}{12}\,;\,\frac{+2}{12},\, \frac{+2}{12},\, \frac{+2}{12}\right),~V_+^2 =\frac{90}{144},\\ [0.4em] 
 V_-=\left( \frac{+1}{12},\, \frac{+1}{12},\,\frac{+1}{12},\,\frac{+1}{12},\,\frac{+1}{12},\,\frac{+1}{12},\,\frac{+1}{12},\,  \frac{+1}{12}, \,\frac{+1}{12};\,  \frac{+11}{12},\, \frac{+6}{12}\,;\, \frac{-2}{12},\, \frac{-2}{12},\,\frac{-2}{12} ,\, \frac{-2}{12},\, \frac{-2}{12} \right),~V_-^2 =\frac{186}{144} %
   \end{array} \label{eq:Model}
\end{eqnarray}
where                                                
\begin{eqnarray}
a_3=a_4=\left(0^9; \,\frac{-2}{3},\,0;\,\frac{2}{3},\,\frac{2}{3},\,\frac{2}{3},\,\frac{2}{3},\,\frac{2}{3}\right). 
\end{eqnarray}
The right-hand side (RHS) sector of the heterotic string is given by the spin lattice  $s=( \ominus~{\rm or}~\oplus;\hat{s})$ with every entry being integer multiples of $\frac12$, satisfying  $s^2=2$.  The  $\hat{\phi}_s$ for $\Z_{12-I}$ with three entries is Eq. (\ref{eq:Z12I}),
  \begin{eqnarray}
  \hat{\phi}_s=\left(\frac{5}{12},\frac{4}{12},\frac{1}{12}\right),~\textrm{with~} \phi_s^2=(\frac{5}{12}, \frac{4}{12}, \frac{1}{12})^2=\frac{42}{144} .
  \label{eq:phishat}
  \end{eqnarray}

\subsection{Untwisted sector $U$} 
\noindent
In $U$, we find the following nonvanishing roots of SU(5)$\times$SU(9)$'\times$U(1)$^4$, 
\begin{eqnarray}
 {\rm SU(5) ~gauge~ multiplet:}~~& P\cdot V =0{\rm~mod.~ integer ~and~}P\cdot a_3=0 {\rm~mod.~ integer}\nonumber\\
{\rm SU(5)}: & \begin{array}{l}  
 P=(0^{9};\,0^2;\, \underline{1\, -1\,0\,0\,0}).
   \end{array}  \label{eq:Blwa3}\\[0.3em]
{ \rm SU(9)'~gauge~ multiplet:}~~& P\cdot V =0{\rm~mod.~ integer}
 \nonumber\\
{\rm SU(9)'}:& \begin{array}{l} P=(\underline{+1\, -1\, 0\, 0\, 0\, 0\, 0\,  0\, 0\,};0\, 0\,;0\, 0\, 0\,  0\, 0).\end{array}
\end{eqnarray}
In addition, there exists U(1)$^4$ symmetry. 
The non-singlet matter fields satisfy
\begin{eqnarray}
  {\rm SU(5)~and/or~SU(9)'~matter~ multiplet:}~&P\cdot V =\frac{1,~4,~5}{12},P\cdot a_3=0 {\rm~mod.~ integer}.
  \label{eq:Untwisted}
\end{eqnarray}
  The conditions (\ref{eq:Untwisted}) allows the  $P^2=2$ lattice shown in Table \ref{tab:Unt}. 
    
 \bigskip 
 \begin{table} [!h]
\begin{center}
\begin{tabular}{|c|c|c| c| }
 \hline  
$U_i$ &\,$P$\, ,&\,Chirality\,&\,$[p_{\rm spin}]~(-p_{\rm spin}\cdot\phi_s)$\, \\[0.1em] \hline  
   ~$U_1\,(p\cdot V=\frac{5}{12})$~  & $(\underline{-1~0^8};0~1 ;0^5)$ & L &$[\ominus;---]~~\left( \frac{+5}{12} \right)$ \\   
    $U_2\,(p\cdot V=\frac{4}{12})$ &--   & &--  \\ 
    $U_3\,(p\cdot V=\frac{1}{12})$  & --&   &--   \\[0.1em]  
\hline
\end{tabular}
\end{center}
\caption{It is a ${\nine}_L $   in view of Eq. (\ref{eq:Vector9}).   }\label{tab:Unt}
\end{table}

\subsection{Twisted sectors}\label{subsec:Twisted}
 \noindent
 In $T_9$, for example, we have
\begin{eqnarray}
9V_0&=&\left( (\frac{+9}{12})^{9}\,;\,\frac{+27}{12}\, \frac{+54}{12}\,;\, (\frac{+54}{12})^5 \, \right), \label{eq:9Tzero}\\ 
9 V_+&=&  \left( (\frac{+9}{12})^{9}\,;\, \frac{-45}{12}\, \frac{+54}{12}\,;\, (\frac{+18}{12})^5\, \right),\label{eq:9Tplus} \\ 
9V_-&=& \left( (\frac{+9}{12})^{9}\,;\, \frac{+99}{12}\, \frac{+54}{12}\,;\, (\frac{-18}{12})^5\, \right).
\label{eq:9Tminus}
\end{eqnarray} 
We shift the lattice by adding SU(9)$'\times$SU(5) singlet vector; $v:(v_1,\cdots,v_{16})$ satisfying $\sum_i |v_i|=0$ modulo $2k$. In the following equations, we list all the shift vectors, $kV_{0,+,-}$, in all twisted sectors. Equations (\ref{eq:9Tzero},\ref{eq:9Tplus},\ref{eq:9Tminus}) are in Eq. (\ref{eq:T9}).

\begin{eqnarray}
&T_1^0\,\left( (\frac{+1}{12})^{9};\,\frac{+3}{12},\frac{+6}{12};(\frac{+6}{12})^5  \right);~ 
T_1^+\,\left( (\frac{+1}{12})^{9};\,\frac{-5}{12},\frac{+6}{12};\, (\frac{+2}{12})^5  \right);~
T_1^-\,\left( (\frac{+1}{12})^{9};\,\frac{+11}{12},\frac{+6}{12};\, (\frac{-2}{12})^5 \right),\label{eq:T2}\\[0.3em] 
&T_2^0\,\left( (\frac{+2}{12})^{9};\,\frac{-6}{12},-1;\, 0^5  \right);~ 
T_2^+\,\left( (\frac{+2}{12})^{9};\,\frac{+2}{12},-1;\, (\frac{-8}{12})^5  \right);~
T_2^-\,\left( (\frac{+2}{12})^{9};\,\frac{-2}{12},-1;\, (\frac{-4}{12})^5 \right),\label{eq:T2}\\[0.3em] 
&T_4^0\, \left( (\frac{+4}{12})^{9};0,0;\,(+1)^5\,\right);~
T_4^+\, \left((\frac{+4}{12})^{9};\frac{-8}{12},\,0;(\frac{-4}{12})^5\,\right);~
T_4^-\,\left((\frac{+4}{12})^{9};\frac{+8}{12},2;(\frac{+4}{12})^5\,\right),\label{eq:T4}\\ [0.3em] 
&T_7^0\, \left( (\frac{-5}{12})^{9};\,\frac{+9}{12},\,\frac{+6}{12};\, (\frac{+6}{12})^5 \right);~ 
T_7^+\,  \left( (\frac{+1}{12})^{9};\,\frac{-35}{12},\,\frac{+42}{12};\, (\frac{+2}{12})^5 \right);~
T_7^-\,  \left( (\frac{+7}{12})^{9};\,\frac{+5}{12},\,\frac{+6}{12};\, (\frac{-2}{12})^5  \right),\label{eq:T7}\\[0.3em] 
&T_9^0\,\left( (\frac{-3}{12})^{9}\,;\,\frac{-9}{12}, \frac{+6}{12}\,;\, (\frac{+6}{12})^5, \right);~
T_9^+\,  \left( (\frac{+9}{12})^{9}\,;\, \frac{+3}{12}, \frac{+6}{12}\,;\, (\frac{-6}{12})^5\, \right);~
T_9^-\, \left( (\frac{+9}{12})^{9}\,;\, \frac{+3}{12}, \frac{-6}{12}\,;\, (\frac{-6}{12})^5\, \right),
\label{eq:T9}\\ [0.3em] 
&T_6^0=\left( (\frac{-6}{12})^9 \,;\,\frac{+18}{12},\,  +3\,;\, 0^5 \right) .   \label{eq:T6}
\end{eqnarray}
 
\subsubsection{Twisted sector $T_9 \,( \delta_3 =\frac{3}{12})$}

\bigskip
\noindent$\bullet$   Two indices spinor-form from $T_9^0$: with Eq. (\ref{eq:9Tzero}) 
 the spinor forms satisfying the mass-shell condition $(P+9V_0)^2=\frac{234}{144}=\frac{13}{8}$  and the Wilson-line condition $12(P+9V_0)\cdot a_3= 0$ are possible for SU(9):
  \begin{equation}
 P_{9}=\left(\underline{\frac{-3}{2}\,\frac{-3}{2}\,-^{7}};\frac{-5}{2},\frac{-9}{2};(\frac{-9}{2})^5 \right).\label{eq:Vtwoindex90}
 \end{equation} 
Instead of the above $P_9$, with Eq. (\ref{eq:T9}) we have
\begin{equation}
P_9=(\underline{--+^7}; +-;-^5)\label{eq:VtwoShift}
\end{equation}
which is a shift from Eq. (\ref{eq:Vtwoindex90}) by $(1^9;3,4;4^5)$. With Eqs.  (\ref{eq:T9})  and  (\ref{eq:VtwoShift}), we present the multiplicity in Table   \ref{tab:AbbTnine}. 
 \begin{table}[!ht]
\begin{center}
\begin{tabular}{|cc|c|cc|ccc|cc|}
 \hline &&&&&&&&& \\[-1.15em]
  Chirality  &   $\tilde s$& $-\tilde{s}\cdot\phi_s$& $-k\,p_{\rm vec}^{ k\,\rm th}\cdot \phi_s$ &$k\,P_{9}\cdot V_0$& $(k/2)\phi_s^2$,&$  -(k/2) V_0^2 $,& ~~$\Delta_{9}^N,~-\delta_9^N
  $   &~~$\Theta_{9},$& Mult. \\[0.15em]
 \hline &&&&&&& && \\[-1.15em]
$\ominus=L$& $(---)$  &  $\frac{+5}{12}$ & $~~\frac{-3}{12},$& $\frac{+6}{12}$ &$\frac{189}{144}  $  &$\frac{+243}{144}$ & $ \frac{0}{12},~~\frac{- 6}{12}$&$\frac{+2}{12} +\frac{-2}{12}  $& 2  \\ [0.1em]
 \hline &&&&&&&&& \\[-1.25em]
  $\oplus=R$& $(+++)$  &  $\frac{-5}{12}$ & $~~\frac{-3}{12}$&$\frac{+6}{12}$  &$\frac{189}{144}$ &$\frac{+243}{144}$&$ \frac{0}{12},~~\frac{-6}{12}$& $\frac{-8}{12}+\frac{-2}{12} $  &1   \\[0.15em] 
\hline
\end{tabular}
\end{center}
\caption{Two index spinor-from $T_9^0:({\one},\tsix')_L$. This is the abbreviated one of Table \ref{tab:ThSix} presented  in Appendix \ref{app:Phihat}. Note that we provided an extra phase $e^{2\pi i(-2/N)}$ as commented below Eq. (\ref{eq:PhaseDiff}).}\label{tab:AbbTnine} 
\end{table}
 
\subsubsection{Twisted sector $T_4\,  ( \delta_4 =0)$}
 
\noindent$\bullet$  One index vector-form  from $T_4^0$:   
\begin{eqnarray}
P_9= (\underline{-1~0^{8}}\,;0~0;(-1)^5), \label{eq:Voneindex40}
\end{eqnarray}
satisfies $(P_9+4V_0)^2=\frac{192}{144}$.   
 Thus, we obtain Table \ref{tab:OneFourZero}.
 
 \begin{table}[!h]
\begin{center}
\begin{tabular}{|cc|c|cc|ccc|cc|}
 \hline &&&&&&&&& \\[-1.15em]
  Chirality  &   $\tilde s$& $-\tilde{s}\cdot\phi_s$& $-k\,p_{\rm vec}^{ k\,\rm th}\cdot \phi_s$ &$k\,P_{9}\cdot V_0$& $(k/2)\phi_s^2$,&$  -(k/2) V_0^2 $,& ~~$\Delta_{4}^N,~-\delta_4^N
  $   &~~$\Theta_{9},$& Mult.  \\[0.15em]
 \hline &&&&&&& && \\[-1.15em]
$\ominus=L$& $(---)$  &  $\frac{+5}{12}$ & $~~\frac{+8}{12},$& $\frac{-4}{12}$ &$\frac{84}{144} $  &$\frac{+108}{144}$ & $ \frac{0}{12},~~\frac{0}{12}$&$\frac{+1}{12}  $& 0\\ [0.1em]
 \hline &&&&&&&&& \\[-1.25em]
  $\oplus=R$& $(+++)$  &  $\frac{-5}{12}$ & $~~\frac{+8}{12}$&$\frac{-4}{12}$ &$\frac{84}{144} $  &$\frac{+108}{144}$ & $ \frac{0}{12},~~\frac{0}{12}$&$\frac{-9}{12}  $&  7 \\[0.15em] 
\hline
\end{tabular}
\end{center}
\caption{One index vector-form from $T_4^0$:  Thus,  we obtain ${  7\cdot ({\one},{\nine}')_R} $. } \label{tab:OneFourZero}    
\end{table}
  
   \bigskip
 \noindent$\bullet$ One index spinor-form for $T_4^-$:  the spinor-form \begin{equation} 
P_9=\left( \underline{+\,-^8};-,\frac{-3}{2};\, {-^5}\right),
\label{eq:VfourPlus}
\end{equation}
 satisfies $(P_9+4V_-)^2=\frac{192}{144}$, 
which is shown in Table \ref{tab:OneFourMinus}.

 \begin{table}[!h]
\begin{center}
\begin{tabular}{|cc|c|cc|ccc|cc|}
 \hline &&&&&&&&& \\[-1.15em]
  Chirality  &   $\tilde s$& $-\tilde{s}\cdot\phi_s$& $-k\,p_{\rm vec}^{ k\,\rm th}\cdot \phi_s$ &$k\,P_{9}\cdot V_-$& $(k/2)\phi_s^2$,&$  -(k/2) V_-^2 $,& ~~$\Delta_{4}^N,~-\delta_4^N
  $   &~~$\Theta_{9},$& Mult.  \\[0.15em]
 \hline &&&&&&& && \\[-1.15em]
$\ominus=L$& $(---)$  &  $\frac{+5}{12}$ & $~~\frac{+8}{12},$& $\frac{-4}{12}$ &$\frac{84}{144} $  &$\frac{-372}{144}$ & $ \frac{0}{12},~~\frac{0}{12}$&$\frac{+9}{12}  $& 7\\ [0.1em]
 \hline &&&&&&&&& \\[-1.25em]
  $\oplus=R$& $(+++)$  &  $\frac{-5}{12}$ & $~~\frac{+8}{12}$&$\frac{-4}{12}$ &$\frac{84}{144} $  &$\frac{-372}{144}$ & $ \frac{0}{12},~~\frac{0}{12}$&$\frac{-1}{12}  $&  0 \\[0.15em] 
\hline
\end{tabular}
\end{center}
\caption{One index spinor-form from $T_4^-$:  Thus,  we obtain ${  7\cdot ({\one},{\nine}')_L} $. } \label{tab:OneFourMinus}    
\end{table}

 \subsubsection{Twisted sector $T_1 \,( \delta_1 =\frac{1}{12})$}
 \noindent
In $T_1$,  we use Eq. (\ref{eq:Model}). 
  
 \bigskip
\noindent$\bullet$  One index vector-form for $T_1^+$: the vector 
\begin{eqnarray}
P_{5}=(0^9;+1,0;\underline{-1,\,0,\,0,\,0,\,0 }), 
\end{eqnarray}
satisfies $(P+V_+)^2=\frac{186}{144}$   which is short by $\frac{24}{144}=\frac{2}{12}$ from the target value of  $\frac{210}{144}$, 
and all states are shown in Table \ref{tab:T1VectPl}.
 
\begin{table}[!ht]
\begin{center}
\begin{tabular}{|cc|c|cc|ccc|cc|}
 \hline &&&&&&&&& \\[-1.15em]
  Chirality  &   $\tilde s$& $-\tilde{s}\cdot\phi_s$& $-k\,p_{\rm vec}^{ k\,\rm th}\cdot \phi_s$ &$k\,P_{5}\cdot V_+$& $(k/2)\phi_s^2$,&$  -(k/2) V_+^2 $,& ~~$\Delta_{k}^N,~-\delta_k^N
  $   &~~$\Theta_5,$& Mult.  \\[0.15em]
 \hline &&&&&&& && \\[-1.15em]
$\ominus=L$& $(---)$  &  $\frac{+5}{12}$ & $~~\frac{+5}{12},$& $\frac{-7}{12}$ &$\frac{+21}{144} $  &$\frac{-45}{144};\frac{-2}{12}$ & $ \frac{+2}{12},~~\frac{-2}{12}$&$\frac{+1}{12}  $& 0 \\ [0.1em]
 \hline &&&&&&&&& \\[-1.25em]
  $\oplus=R$& $(+++)$  &  $\frac{-5}{12}$ & $~~\frac{+5}{12}$&$\frac{-8}{12}$ &$\frac{+21}{144} $  &$\frac{-45}{144}$ & $ \frac{+2}{12},~~\frac{-2}{12}$&$\frac{-9}{12}  $&  3  \\[0.15em] 
\hline
\end{tabular}
\end{center}
\caption{One index vector-form from $T_1^+$:   Thus, we obtain $ 3\cdot({\five},{\one})_R  $ that is $\rho$ in Table \ref{tab:Quints}. } \label{tab:T1VectPl}
\end{table}

\begin{table}[!ht]
\begin{center}
\begin{tabular}{|cc|c|cc|ccc|cc|}
 \hline &&&&&&&&& \\[-1.15em]
  Chirality  &   $\tilde s$& $-\tilde{s}\cdot\phi_s$& $-k\,p_{\rm vec}^{ k\,\rm th}\cdot \phi_s$ &$k\,P_{5}\cdot V_+$& $(k/2)\phi_s^2$,&$  -(k/2) V_+^2 $,& ~~$\Delta_{k}^N,~-\delta_k^N
  $   &~~$\Theta_5,$& Mult.  \\[0.15em]
 \hline &&&&&&& && \\[-1.15em]
$\ominus=L$& $(---)$  &  $\frac{+5}{12}$ & $~~\frac{+5}{12},$& $\frac{-8}{12}$ &$\frac{+21}{144} $  &$\frac{-45}{144};\frac{-2}{12}$ & $ \frac{+2}{12},~~\frac{-2}{12}$&$\frac{0}{12}  $& 3 \\ [0.1em]
 \hline &&&&&&&&& \\[-1.25em]
  $\oplus=R$& $(+++)$  &  $\frac{-5}{12}$ & $~~\frac{+5}{12}$&$\frac{-8}{12}$ &$\frac{+21}{144} $  &$\frac{-45}{144}$ & $ \frac{+2}{12},~~\frac{-2}{12}$&$\frac{-10}{12}  $&  0  \\[0.15em] 
\hline
\end{tabular}
\end{center}
\caption{One index vector-form from $T_1^+$:   Thus, we obtain $ 3\cdot({\five},{\one})_L  $ that is $\sigma$ in Table \ref{tab:Quints}. } \label{tab:T1VectPlus}
\end{table}

 \bigskip
\noindent$\bullet$  One index vector-form for $T_1^+$: the vector 
\begin{eqnarray}
P_{5}=(0^9;0,-1;\underline{-1,\,0,\,0,\,0,\,0 }), 
\end{eqnarray}
satisfies $(P+V_+)^2=\frac{186}{144}$   which is short by $\frac{24}{144}=\frac{2}{12}$ from the target value of  $\frac{210}{144}$, 
and all states are shown in Table \ref{tab:T1VectPlus}.

 \bigskip
\noindent$\bullet$  One index vector-form for $T_1^-$: the vector 
\begin{eqnarray}
P_{9}=(\underline{-1~0^8};-1,0;0^5), 
\end{eqnarray}
satisfies $(P+V_-)^2=\frac{186}{144}$   which is short by $\frac{24}{144}=\frac{2}{12}$ from the target value of  $\frac{210}{144}$, and the spectrum is shown in Table \ref{tab:T1VectorMinus}.

\begin{table}[!ht]
\begin{center}
\begin{tabular}{|cc|c|cc|ccc|cc|}
 \hline &&&&&&&&& \\[-1.15em]
  Chirality  &   $\tilde s$& $-\tilde{s}\cdot\phi_s$& $-k\,p_{\rm vec}^{ k\,\rm th}\cdot \phi_s$ &$k\,P_{5}\cdot V_-$& $(k/2)\phi_s^2$,&$  -(k/2) V_-^2 $,& ~~$\Delta_{k}^N,~-\delta_k^N
  $   &~~$\Theta_5,$& Mult.  \\[0.15em]
 \hline &&&&&&& && \\[-1.15em]
$\ominus=L$& $(---)$  &  $\frac{+5}{12}$ & $~~\frac{+5}{12},$& $\frac{0}{12}$ &$\frac{+21}{144} $  &$\frac{-93}{144}$ & $ \frac{+2}{12},~~\frac{-2}{12}$&$\frac{+4}{12}  $&0\\ [0.1em]
 \hline &&&&&&&&& \\[-1.25em]
  $\oplus=R$& $(+++)$  &  $\frac{-5}{12}$ & $~~\frac{+5}{12}$&$\frac{0}{12}$ &$\frac{+21}{144} $  &$\frac{-93}{144}$ & $ \frac{+2}{12},~~\frac{-2}{12}$&$\frac{-6}{12}  $& 3 \\[0.15em] 
\hline
\end{tabular}
\end{center}
\caption{One index vector-form from $T_1^-$:   Thus, we obtain $ 3\cdot({\one} , \nine')_R $. } \label{tab:T1VectorMinus}
\end{table}
 
 \bigskip
\noindent$\bullet$  One index vector-form from $T_1^-$: the vector 
\begin{eqnarray}
P_{5}=(0^9;-1,0;\underline{+1\,0^4 }), 
\end{eqnarray}
satisfies $(P+V_-)^2=\frac{162}{144}$   which is short by $\frac{4}{12}$ from the target value of  $\frac{210}{144}$,  
 and we obtain Table  \ref{tab:T1VectorPlus}. 

\begin{table}[!ht]
\begin{center}
\begin{tabular}{|cc|c|cc|ccc|cc|}
 \hline &&&&&&&&& \\[-1.15em]
  Chirality  &   $\tilde s$& $-\tilde{s}\cdot\phi_s$& $-k\,p_{\rm vec}^{ k\,\rm th}\cdot \phi_s$ &$k\,P_{5}\cdot V_-$& $(k/2)\phi_s^2$,&$  -(k/2) V_-^2 $,& ~~$\Delta_{k}^N,~-\delta_k^N
  $   &~~$\Theta_5,$& Mult.  \\[0.15em]
 \hline &&&&&&& && \\[-1.15em]
$\ominus=L$& $(---)$  &  $\frac{+5}{12}$ & $~~\frac{+5}{12},$& $\frac{-1}{12}$ &$\frac{+21}{144} $  &$\frac{-93}{144}$ & $ \frac{+4}{12},~~\frac{-2}{12}$&$\frac{+5}{12}  $& 3 \\ [0.1em]
 \hline &&&&&&&&& \\[-1.25em]
  $\oplus=R$& $(+++)$  &  $\frac{-5}{12}$ & $~~\frac{+5}{12}$&$\frac{-1}{12}$ &$\frac{+21}{144} $  &$\frac{-93}{144}$ & $ \frac{+4}{12},~~\frac{-2}{12}$&$\frac{-5}{12}  $&  3  \\[0.15em] 
\hline
\end{tabular}
\end{center}
\caption{One index vector-form from $T_1^-$:   Thus, we obtain $ 3\cdot( \five,{\one})_L+3\cdot(\five,{\one})_R $ that are $\beta$ and $\alpha$  in Table \ref{tab:Quints}. . } \label{tab:T1VectorPlus}
\end{table}

\subsubsection{Twisted sector $T_2 \,( \delta_2 =\frac{0}{12} )$}

\bigskip
\noindent$\bullet$ One index vector-form from $T_2^0$:
 the vector 
\begin{eqnarray}
P_{9}=(\underline{-1\, 0^8};0,+1;0^5), 
\end{eqnarray}
which gives $(P+2V_+)^2=\frac{168}{144}$ which is short by $\frac{48}{144}=\frac{4}{12}$ from the target value of $\frac{216}{144}$, and massless fields are shown in Table \ref{tab:T2oneZero}.
\begin{table}[!ht]
\begin{center}
\begin{tabular}{|cc|c|cc|ccc|cc|}
 \hline &&&&&&&&& \\[-1.15em]
  Chirality  &   $\tilde s$& $-\tilde{s}\cdot\phi_s$& $-k\,p_{\rm vec}^{ k\,\rm th}\cdot \phi_s$ &$k\,P \cdot V_0$& $(k/2)\phi_s^2$,&$  -(k/2) V_0^2 $,& ~~$\Delta_{2}^N,~-\delta_2^N
  $   &~~$\Theta_{5},$& Mult. \\[0.15em]
 \hline &&&&&&& && \\[-1.15em]
$\ominus=L$& $(---)$  &  $\frac{+5}{12}$ & $~~\frac{+10}{12},$& $\frac{-2}{12}$ &$\frac{+42}{144} $  &$\frac{+54}{144};\frac{8}{12}$ & $ \frac{+4}{12},~~\frac{0}{12}$&$\frac{+1}{12}  $& 0\\ [0.1em]
 \hline &&&&&&&&& \\[-1.25em]
  $\oplus=R$& $(+++)$  &  $\frac{-5}{12}$ & $~~\frac{+10}{12}$&$\frac{-2}{12}$ &$\frac{+42}{144} $  &$\frac{+54}{144}$ & $ \frac{+4}{12},~~\frac{0}{12}$&$\frac{-9}{12}  $&  3 \\[0.15em] 
\hline
\end{tabular}
\end{center}
\caption{One index vector-form from $T_2^0$:  Thus, we obtain ${3\cdot ({\one},{\nine}')_R} $.  } \label{tab:T2oneZero} 
\end{table}

\bigskip
\noindent$\bullet$ One index spinor-form from $T_2^-$:
 the vector 
\begin{eqnarray}
P_{5}=(-^9;+,+;\underline{-++++}), 
\end{eqnarray}
which gives $(P+2V_-)^2=\frac{192}{144}$ which is short by $ \frac{2}{12}$ from the target value of $\frac{216}{144}$, and massless fields are shown in Table \ref{tab:TtwoMinus}  .
  
\begin{table}[!ht]
\begin{center}
\begin{tabular}{|cc|c|cc|ccc|cc|}
 \hline &&&&&&&&& \\[-1.15em]
  Chirality  &   $\tilde s$& $-\tilde{s}\cdot\phi_s$& $-k\,p_{\rm vec}^{ k\,\rm th}\cdot \phi_s$ &$k\,P \cdot V_-$& $(k/2)\phi_s^2$,&$  -(k/2) V_-^2 $,& ~~$\Delta_{2}^N,~-\delta_2^N
  $   &~~$\Theta_{5},$& Mult. \\[0.15em]
 \hline &&&&&&& && \\[-1.15em]
$\ominus=L$& $(---)$  &  $\frac{+5}{12}$ & $~~\frac{+10}{12},$& $\frac{+4}{12}$ &$\frac{+42}{144} $  &$\frac{-90}{144}; \frac{-4}{12}$ & $ \frac{+2}{12},~~\frac{0}{12}$&$\frac{+5}{12}  $& 3\\ [0.1em]
 \hline &&&&&&&&& \\[-1.25em]
  $\oplus=R$& $(+++)$  &  $\frac{-5}{12}$ & $~~\frac{+10}{12}$&$\frac{+4}{12}$ &$\frac{+42}{144} $  &$\frac{-90}{144}$ & $ \frac{+2}{12},~~\frac{0}{12}$&$\frac{-5}{12}  $&  3 \\[0.15em] 
\hline
\end{tabular}
\end{center}
\caption{One index spinor-form from $T_2^-$:  Thus, we obtain $3\cdot ({\fiveb},{\one})_L+3\cdot ({\fiveb},{\one})_R $ that are $\eta$ and  $\xi$ in Table \ref{tab:Quints}.} \label{tab:TtwoMinus} 
\end{table}

\subsubsection{Twisted sector $T_7 \,( \delta_7 =\frac{1}{12} )$}
  
\noindent$\bullet$  One index spinor-form from $T_7^0$: 
\begin{eqnarray}
P_{9}=\left(\underline{-\,+^{8}};-,-;-^5\right), \label{eq:V7Zero2s
}
\end{eqnarray}
gives $(P_9+7V_0)^2=  \frac{138}{144}$, which is short of $\frac{6}{12}$ from $\frac{210}{12}$, and we obtain Table  \ref{tab:T7zero9}.

\begin{table}[!ht]
\begin{center}
\begin{tabular}{|cc|c|cc|ccc|cc|}
 \hline &&&&&&&&& \\[-1.15em]
  Chirality  &   $\tilde s$& $-\tilde{s}\cdot\phi_s$& $-k\,p_{\rm vec}^{ k\,\rm th}\cdot \phi_s$ &$k\,P_{9}\cdot V_0$& $(k/2)\phi_s^2$,&$  -(k/2) V_0^2 $,& ~~$\Delta_{k}^N,~-\delta_k^N
  $   &~~$\Theta_{9},$& Mult. \\[0.15em]
 \hline &&&&&&& && \\[-1.15em]
$\ominus=L$& $(---)$  &  $\frac{+5}{12}$ & $~~\frac{+1}{12},$& $\frac{-4}{12}$ &$\frac{+147}{144};\frac{4}{12} $  &$\frac{+189}{144}$ & $ \frac{+6}{12},~~\frac{-2}{12}$&$\frac{+10}{12}  $& 0 \\ [0.1em]
 \hline &&&&&&&&& \\[-1.25em]
  $\oplus=R$& $(+++)$  &  $\frac{-5}{12}$ & $~~\frac{+1}{12}$&$\frac{-4}{12}$ &$\frac{+147}{144} $  &$\frac{+189}{144}$ & $ \frac{+6}{12},~~\frac{-2}{12}$&$\frac{0}{12}  $&  3 \\[0.15em] 
\hline
\end{tabular}
\end{center}
\caption{One index spinor-form from $T_7^0$: $3\cdot ({\one},{\nine}')_R  $.}\label{tab:T7zero9}  
\end{table}

\begin{table}[!ht]
\begin{center}
\begin{tabular}{|cc|c|cc|ccc|cc|}
 \hline &&&&&&&&& \\[-1.15em]
  Chirality  &   $\tilde s$& $-\tilde{s}\cdot\phi_s$& $-k\,p_{\rm vec}^{ k\,\rm th}\cdot \phi_s$ &$k\,P_{9}\cdot V_+$& $(k/2)\phi_s^2$,&$  -(k/2) V_+^2 $,& ~~$\Delta_{k}^N,~-\delta_k^N
  $   &~~$\Theta_{9},$& Mult.\\[0.15em]
 \hline &&&&&&& && \\[-1.15em]
$\ominus=L$& $(---)$  &  $\frac{+5}{12}$ & $~~\frac{+1}{12},$& $\frac{+2}{12}$ &$\frac{+147}{144} $  &$\frac{-315}{144};\frac{-2}{12}$ & $ \frac{+2}{12},~~\frac{-2}{12}$&$\frac{+6}{12}  $& 3 \\ [0.1em]
 \hline &&&&&&&&& \\[-1.25em]
  $\oplus=R$& $(+++)$  &  $\frac{-5}{12}$ & $~~\frac{+1}{12}$&$\frac{+2}{12}$ &$\frac{+147}{144} $  &$\frac{-315}{144}$ & $ \frac{+2}{12},~~\frac{-2}{12}$&$\frac{-4}{12}  $&  0 \\[0.15em] 
\hline
\end{tabular}
\end{center}
\caption{One index vector-form from $T_7^+$: $ 3\cdot({\one},{\nine}')_L$.  }  
\label{tab:TOnePlus}
\end{table}
 \bigskip
\noindent $\bullet$  Two index vector-form from $T_7^+$:
\begin{eqnarray}
P_{9}=\left(\underline{-1~0^8};+3,-4;0^5\right),  \label{eq:T5Minus1}
\end{eqnarray}
gives $(P_9+7V_+)^2=  \frac{186}{144}$ which is short of $\frac{2}{12}$ from the 
target value $\frac{210}{144}$, and massless fields are shown in Table \ref{tab:TOnePlus}.

 \bigskip
\noindent$\bullet$  Two index spinor-form from $T_7^-$: The vector
 \begin{eqnarray}
P_{5}=\left(-^9;-\,-;\underline{+\,+\,+\,-\,-} \right),  \label{eq:T5Minus1}
\end{eqnarray}
gives $(P_5+7V_-)^2=  \frac{186}{144}$  which is short of $\frac{2}{12}$ from the  target value $\frac{210}{144}$, and massless fields are shown in Table \ref{tab:TsvMinTwo}.

\begin{table}[!h]
\begin{center}
\begin{tabular}{|cc|c|cc|ccc|cc|}
 \hline &&&&&&&&& \\[-1.15em]
  Chirality  &   $\tilde s$& $-\tilde{s}\cdot\phi_s$& $-k\,p_{\rm vec}^{ k\,\rm th}\cdot \phi_s$ &$k\,P_{5}\cdot V_-$& $(k/2)\phi_s^2$,&$  -(k/2) V_-^2 $,& ~~$\Delta_{k}^N,~-\delta_k^N
  $   &~~$\Theta_{5},$& Mult.  \\[0.15em]
 \hline &&&&&&& && \\[-1.15em]
$\ominus=L$& $(---)$  &  $\frac{+5}{12}$ & $~~\frac{+1}{12},$& $\frac{-2}{12}$ &$\frac{+147}{144} $  &$\frac{-651}{144};\frac{6}{12}$ & $ \frac{+2}{12},~~\frac{-2}{12}$&$\frac{+10}{12}  $& 0 \\ [0.1em]
 \hline &&&&&&&&& \\[-1.25em]
  $\oplus=R$& $(+++)$  &  $\frac{-5}{12}$ & $~~\frac{+1}{12}$&$\frac{-2}{12}$ &$\frac{+147}{144} $  &$\frac{-651}{144}$ & $ \frac{+2}{12},~~\frac{-2}{12}$&$\frac{0}{12}  $&  3 \\[0.15em] 
\hline
\end{tabular}
\end{center}
\caption{Two index spinor-form from $T_7^-$: $3\,({\tenb},{\one})_R$  that is $T$ in Table \ref{tab:Quints}.}  
\label{tab:TsvMinTwo}
\end{table}

 \bigskip
\noindent$\bullet$  One index spinor-form from $T_7^-$: The vector
 \begin{eqnarray}
P_{5}=\left(-^9;-\,-;\underline{+\,+\,+\,+\,-} \right),  \label{eq:T5Minus1}
\end{eqnarray}
gives $(P_5+7V_-)^2=  \frac{138}{144}$  which is short of $\frac{6}{12}$ from the  target value $\frac{210}{144}$, and massless fields are shown in Table \ref{tab:TsvMinOne}.

\begin{table}[!ht]
\begin{center}
\begin{tabular}{|cc|c|cc|ccc|cc|}
 \hline &&&&&&&&& \\[-1.15em]
  Chirality  &   $\tilde s$& $-\tilde{s}\cdot\phi_s$& $-k\,p_{\rm vec}^{ k\,\rm th}\cdot \phi_s$ &$k\,P_{5}\cdot V_-$& $(k/2)\phi_s^2$,&$  -(k/2) V_-^2 $,& ~~$\Delta_{k}^N,~-\delta_k^N
  $   &~~$\Theta_{5},$& Mult.  \\[0.15em]
 \hline &&&&&&& && \\[-1.15em]
$\ominus=L$& $(---)$  &  $\frac{+5}{12}$ & $~~\frac{+1}{12},$& $\frac{-4}{12}$ &$\frac{+147}{144} $  &$\frac{-651}{144}$ & $ \frac{+6}{12},~~\frac{-2}{12}$&$\frac{0}{12}  $& 3 \\ [0.1em]
 \hline &&&&&&&&& \\[-1.25em]
  $\oplus=R$& $(+++)$  &  $\frac{-5}{12}$ & $~~\frac{+1}{12}$&$\frac{-4}{12}$ &$\frac{+147}{144} $  &$\frac{-651}{144}$ & $ \frac{+6}{12},~~\frac{-2}{12}$&$\frac{-10}{12}  $&  0 \\[0.15em] 
\hline
\end{tabular}
\end{center}
\caption{One index spinor-form from $T_7^-$: $3\,({\fiveb},{\one})_L$  that is $\overline{F}$ in Table \ref{tab:Quints}.}  
\label{tab:TsvMinOne}
\end{table}

\subsubsection{Twisted sector $T_6 \,( \delta_6 =\frac{0}{12} )$}
\label{Ssec:Tsix}

\noindent
Note that $\frac{k}{2}(V_a^2-\phi_s^2)$ in Eq. (\ref{eq:Phase}) is not distinguished by Wilson lines, and  we just calculate the spectra from $6V_0$ with multiplicity  3 (for  $T_6^{0,+,-}$).  With 3 and $0^5$ in (\ref{eq:T6}), we obtain only vectorlike representations. To clarify, we list all the possibilities for $\five,\fiveb,\ten,$ and $\tenb$. All the allowed ones are spinor forms.
 
 \begin{table}[!ht]
\begin{center}
\begin{tabular}{|cc|c|cc|ccc|cc|}
 \hline &&&&&&&&& \\[-1.15em]
  Chirality  &   $\tilde s$& $-\tilde{s}\cdot\phi_s$& $-k\,p_{\rm vec}^{ k\,\rm th}\cdot \phi_s$ &$k\,P_{1}\cdot V_0$& $(k/2)\phi_s^2$,&$  -(k/2) V_0^2 $,& ~~$\Delta_{6}^N,~-\delta_6^N
  $   &~~$\Theta_{5},$& Mult. of $\ten$ and $\tenb$s.\\[0.15em]
 \hline &&&&&&& && \\[-1.15em]
$\ominus=L$& $(---)$  &  $\frac{+5}{12}$ & $~~\frac{0}{12},$& $\frac{0}{12}$ &$\frac{+126}{144}  $  &$\frac{+162}{144}$ & $ \frac{0}{12},~~\frac{0}{12}$&$\frac{+5}{12}$& 6 \\ [0.1em]
 \hline &&&&&&&&& \\[-1.25em]
  $\oplus=R$& $(+++)$  &  $\frac{-5}{12}$ & $~~\frac{0}{12}$&$\frac{0}{12}$  &$\frac{+126}{144}$ &$\frac{+162}{144}$&$ \frac{0}{12},~~\frac{0}{12}$& $\frac{-5}{12}$  &6 \\[0.15em] 
\hline
\end{tabular}
\end{center}
\caption{ $T_6$:  $\ten$  appears 6 times for L and 6 times for R. The same multiplicities occur also in $T_6^{+,-}$. So, in total, there appear 36 multiplicities of $\ten$ for L and also for R.} \label{tab:TSixFive} 
\end{table}

 \bigskip
\noindent$\bullet$ Spinor-form for $\ten$, 
\begin{equation}
P_5=(+^9; \frac{-3}{2},  \frac{-5}{2} {\rm ~or~} \frac{-7}{2};\underline{++\,---})  \label{eq:ten1}
\end{equation}
which saturates the masslessness condition $(P_5+6V_0)^2=\frac{216}{144}$. The spectra are shown in Table \ref{tab:TSixFive}.  Considering Eq. (\ref{eq:ten1}) for $\frac{-5}{2}$ and  $\frac{-7}{2}$, we obtain 12 each for L and R.  Thus, we obtain $36\cdot (\ten_L\oplus \ten_R)$.

\bigskip
\noindent$\bullet$ Spinor-form for $\tenb$, 
\begin{equation}
P_5=(+^9; \frac{-3}{2},  \frac{-5}{2} {\rm ~or~} \frac{-7}{2};\underline{+++\,--})  \label{eq:five1}
\end{equation}
which saturates the masslessness condition $(P_5+6V_0)^2=\frac{216}{144}$. So, we obtain $36\cdot (\tenb_L\oplus \tenb_R)$.

 \bigskip
\noindent$\bullet$ Spinor-form for $\five$, 
\begin{equation}
P_5=(+^9; \frac{-3}{2},  \frac{-5}{2} {\rm ~or~} \frac{-7}{2};\underline{+\,----})  \label{eq:five1}
\end{equation}
which saturates the masslessness condition $(P_5+6V_0)^2=\frac{216}{144}$, and we obtain $36\cdot (\five_L\oplus \five_R)$. 

 \bigskip
\noindent$\bullet$ Spinor-form for $\fiveb$, 
\begin{equation}
P_5=(+^9; \frac{-3}{2},  \frac{-5}{2} {\rm ~or~} \frac{-7}{2};\underline{++++\,-})  \label{eq:five1}
\end{equation}
which saturates the masslessness condition $(P_5+6V_0)^2=\frac{216}{144}$, and we obtain $36\cdot (\fiveb_L\oplus \fiveb_R)$. 

\bigskip
\noindent
Summarizing the non-singlet chiral fields 
we obtained, $ \nine'_L(U)+\tsix'_L(T_9^0)+(\tsix'_L(T_9^0)+\tsix'_R(T_9^0))  +7\cdot \nine'_R(T_4^0)+7\cdot \nine'_L(T_4^-)+3\cdot \nine'_R(T_1^-) +   3\cdot \nine'_R(T_2^0 ) \color{black}+   3\cdot \nine'_R(T_7^0 )+   3\cdot \nine'_L(T_7^+ )$ for SU(9)$'$. After removing vector-like representations, note that   the SU(9)$'$ anomaly is absent with
\begin{eqnarray}
\tsixb'_R+5\cdot \nine'_R. \label{eq:FNine}
 \end{eqnarray}
Similarly after removing vector-like representations, for SU(5) we obtain
 \begin{eqnarray}
3\cdot  {\ten}_L(T_7^+)  +3\cdot  {\fiveb}_L(T_7^-)  .\label{eq:FFive}
\end{eqnarray}
 These spectra in Eqs. (\ref{eq:FNine}) and (\ref{eq:FFive}) do not lead to non-Abelian gauge anomalies.   For $\fiveb_L$ in Eq. (\ref{eq:FFive}), in fact it is a linear combination of $\fiveb_L(T_7^-)$ ($\overline{F}$ in Table \ref{tab:Quints}) and $\fiveb_L(T_2^-)$ ($\eta$ in Table \ref{tab:Quints}). In addition, note that there appear 72 pairs of $\ten\oplus\tenb$ from $T_6$, which are needed for breaking the anti-SU(5) GUT.

\subsection{Singlets, quintets, and Higgs fields}\label{subsec:Higgs}

\noindent 
We obtained the rank 16 group   SU(5)$\times$SU(9)$'\times$U(1)$^4$. The four U(1) charges are also shown in the following tables,  where $Q_X$ is the U(1)$_X$ charge of \anti~GUT.   
The four U(1) charges in the gauge group SU(9)$'\times$SU(5)$\times$U(1)$^4$ are defined by
\begin{eqnarray}
&&Q_1 =\frac19(-2,-2,-2,-2,-2,-2,-2,-2,-2;0,0; 0^5)\label{eq:Q1}\\
&&Q_2=(0^9;-2,0;0^5)\label{eq:Q2}\\
&&Q_3=(0^9;0, -2;0^5)\label{eq:Q3}\\
&&Q_{X}=(0^9;0,0;-2,-2,-2,-2,-2) .\label{eq:QX}
\end{eqnarray} 
 We summarize the  SU(5)$\times$SU(9)$'$ singlets in the following Tables \ref{tab:SingOr}, \ref{tab:ChiralSinglets} and  \ref{tab:SingR},  with the format of Subsec. \ref{subsec:Twisted}. In Table \ref{tab:SingR}, we tabulate the four U(1) charges given in Eqs. (\ref{eq:Q1})--(\ref{eq:QX}). For the SU(5) singlets, the electromagnetic charge is
\begin{equation}
\Qem= \frac15Q_X . 
\end{equation}

 \begin{table}[!ht]
\begin{center}
\begin{tabular}{|c|c|c|c|cc|c|c|c|}
 \hline &&&&&&& \\[-1.15em]
 Model($T_k^{0,+,-}$)  & $-\tilde{s}\cdot\phi_s$& $-k\,p_{\rm vec}^{ k\,\rm th}\cdot \phi_s$ &$k\,P_{\rm Group}\cdot V_{0,+,-}$& $(k/2)\phi_s^2$,&$  -(k/2) V_{0,+,-}^2 $& ~~$\Delta_{\rm Group}^N,~-\delta_{\rm Group}^N
  $   &~~$\Theta_{\rm Group}$ \\[0.15em]
 \hline &&&&&& &  \\[-1.15em]
 $A\,(T_9^0)$ &  $\frac{+5}{12}$ & $~~\frac{-3}{12},$& $\frac{0}{12}$ &$\frac{189}{144}  $  &$\frac{+243}{144} $ & $ \frac{0}{12},\qquad~~\frac{- 6}{12}$&$\frac{-4}{12}   $  \\ [0.1em]
 &&&&&&&  \\[-1.15em]
$B\,(T_9^+) $&  $\frac{+5}{12}$ &$~~\frac{-3}{12},$& $\frac{0}{12}$  &$\frac{189}{144} $  &$\frac{-405}{144}$& $ \frac{0}{12},\qquad~~\frac{-6}{12}$&$\frac{+2}{12}  $ \\[0.15em]  
  &&&&&&&  \\[-1.25em]
$C\,(T_9^-)$ &  $\frac{+5}{12}$ &$~~\frac{-3}{12},$& $\frac{0}{12}$  &$\frac{189}{144} $  &$\frac{-837}{144}$& $ \frac{0}{12},\qquad~~\frac{-6}{12}$&$\frac{+2}{12}  $ \\[0.15em]  
 \hline  &&&&&&&  \\[-1.25em]
$D\,(T_4^+)$ & $\frac{+5}{12}$ &$~~\frac{+8}{12}$&$\frac{+4}{12}$  &$\frac{84}{144} $  &$\frac{-180}{144}$&$ \frac{+8}{12},\qquad~~\frac{0}{12}$& $\frac{+5}{12}  $   \\[0.15em]  
 &&&&&&&  \\[-1.25em]
 $E\,(T_4^-)$  &$\frac{+5}{12}$&$~~\frac{+8}{12}$& $\frac{+4}{12}$  &$\frac{84}{144} $  &$\frac{-372}{144} $& $ \frac{+8}{12},\qquad~~\frac{0}{12}$&$\frac{+1}{12}$\\[0.1em]  
 \hline &&&&&&&  \\[-1.25em]
  $F(T_1^-)$ &  $\frac{+5}{12}$ & $~~\frac{+5}{12}$ & 
$\frac{-5}{12}$  & $\frac{21}{144}$ & $\frac{-93}{144}$ & $ \frac{0}{12},\qquad~~\frac{-2}{12}$ & $\frac{+9}{12}$  \\ [0.1em]
 \hline &&&&&&& \\[-1.25em]
$G(T_2^0)$ &  $\frac{+5}{12}$ & $~~\frac{+10}{12}$ & $\frac{-6}{12}$ & $\frac{42}{144} $  & $\frac{+54}{144}$ &$ \frac{+8}{12},\qquad~~\frac{0}{12}$& $\frac{+1}{12}  $ \\[0.15em] 
 &&&&&&&  \\[-1.25em]
$H(T_2^-)$ &  $\frac{+5}{12}$ & $~~\frac{+10}{12}$ & $\frac{-2}{12}$ & $\frac{42}{144} $  & $\frac{-186}{144} $ &$ \frac{0}{12},\qquad~~\frac{0}{12}$& $\frac{+1}{12}  $ \\[0.15em] 
\hline &&&&&&&  \\[-1.25em]
  $I\,(T_7^0)$  &$\frac{+5}{12}$&$~~\frac{+1}{12}$& $\frac{-6}{12}$  &$\frac{147}{144} $  &$\frac{+189}{144} $& $ \frac{+10}{12},\qquad~~\frac{-2}{12}$&$\frac{0}{12} $  \\[0.1em]  
 &&&&&&&\\[-1.25em]
$J(T_7^-)$  &  $\frac{+5}{12}$ & $~~\frac{+1}{12}$ & 
$\frac{-1}{12}$  & $\frac{147}{144}$ & $\frac{-651}{144} $ & $ \frac{0}{12},\qquad~~\frac{-2}{12}$ & $\frac{+9}{12}$ \\ [0.1em]
 \hline 
\end{tabular}
\end{center}
\caption{Entries in calculating the multiplicities in the following Table \ref{tab:ChiralSinglets}. }\label{tab:SingOr}  
\end{table}

 \begin{table}[!ht]
\begin{center}
\begin{tabular}{|c|c|c|c|c|}
 \hline &&&&  \\[-1.15em]
  Twisted Sector &   $P~;~P+kV_i$& $-\hat{s}\cdot\phi_s$
    &   $\Theta_k$ &Mult. of singlets (L or R)  \\
 \hline &&&& \\[-1.4em]
$A(T_9^0)$& $\left(+^9;+,-;-^5\right)$  & $\frac{5}{12}$& $\frac{-4}{12}$& 1(L) \\ 
 &&&&  \\[-1.25em]
$B(T_9^+)$& $\left(-^9;-,-;+^5\right)$  & $\frac{5}{12}$& $\frac{+2}{12}$&  1(R) \\ 
 &&&&  \\[-1.25em] 
$C(T_9^-)$& $\left(-^9;-,+;+^5\right)$  & $\frac{5}{12}$& $\frac{+2}{12}$&  1(R) \\ 
 &&&&  \\[-1.25em]
\hline &&&& \\[-1.15em]
$D(T_4^+)$& $\left(-^9;+,\pm;+^5\right)$   & $\frac{5}{12}$ &  $\frac{+5}{12}$& 3(L)+3(L), 3(R)+3(R) \\ [0.1em]
 && && \\[-1.25em]
$E(T_4^-)$& $\left(-^9;-,\frac{-3}{2}{\rm ~or~}\frac{-5}{2};-^5\right)$  & $\frac{5}{12}$ &  $\frac{+1}{12}$& 7(R)+7(R)  \\ [0.1em]
\hline && && \\[-1.25em]
 $F(T_1^-)$& $\left(0^9;-1,-1;0^5\right)$  & $\frac{5}{12}$ &  $\frac{+9}{12}$& 3(L),~neutral\\ [0.15em]
 \hline
$G(T_2^0)$& $\left(0^9;+1,+1;0^5\right)$  & $\frac{5}{12}$ &  $\frac{+1}{12}$&   3(R),~neutral \\ [0.15em]
   &&&&  \\[-1.25em]
$H(T_2^-)$& $\left(0^9;+1,+1;0^5\right)$  & $\frac{5}{12}$ &  $\frac{+1}{12}$&   3(R),~neutral \\ [0.15em]
  \hline  &&&&  \\[-1.25em]
  $I\,(T_7^0)$ & $\left(+^9;\frac{-3}{2},-;-^5\right)$  & $\frac{5}{12}$ &  $\frac{0}{12}$& 3(L) \\ [0.25em]
$J(T_7^-)$& $ \left(-^9;+,-;+^5\right)$  & $\frac{5}{12}$ &  $\frac{+9}{12}$& 3(L) \\ [0.25em]
\hline
$K_1(T_6^0)$& $\left( +^9;\frac{-3}{2},\frac{-5}{2};+^5\right)$ & $\frac{5}{12}$  &  $\frac{+5}{12}$ & $3\times 7(L),~3 \times 7(R)  $ \\[0.25em] 
$K_2(T_6^0)$& $\left( +^9;\frac{-3}{2},\frac{-7}{2};+^5\right)$ & $\frac{5}{12}$  &  $\frac{+5}{12}$ & $3\times 7(L),~3 \times 7(R)  $ \\[0.25em] 
$K_3(T_6^0)$& $\left( +^9;\frac{-3}{2},\frac{-5}{2};-^5\right)$ & $\frac{5}{12}$  &  $\frac{+5}{12}$ & $3\times 7(L),~3 \times 7(R)  $ \\[0.25em] 
$K_4(T_6^0)$& $\left( +^9;\frac{-3}{2},\frac{-7}{2};-^5\right)$ & $\frac{5}{12}$  &  $\frac{+5}{12}$ & $3\times 7(L),~3 \times 7(R)  $ \\[0.25em]  
\hline\hline
$T_6 (\ten)$& $(+^9; \frac{-3}{2},  \frac{-5}{2} {\rm ~or~} \frac{-7}{2};\underline{++\,---})$  & $\frac{5}{12}$ &  $\frac{+5}{12}$& $36(L),~36(R)  $ \\ [0.35em]
$T_6 (\tenb)$& $(+^9; \frac{-3}{2}, \frac{-5}{2} {\rm ~or~} \frac{-7}{2};\underline{+++\,--}) $  & $\frac{5}{12}$ &  $\frac{+5}{12}$& $36(L),~36(R)  $  \\ [0.35em]
\hline
\end{tabular}
\end{center}
\caption{Summary of chiral singlets, $\one$'s. The value for $-\hat{s}\cdot\phi_s$ is for $(\ominus;---)$, \ie the ones in the top rows in the tables. We also listed the vectorlike $({\ten}\oplus  {\tenb}, {\one})_R $ obtained previously in Table \ref{tab:TSixFive}.}\label{tab:ChiralSinglets} 
\end{table}

 \begin{table}[!ht]
\begin{center}
\begin{tabular}{|c|c|ccc|cc|c|}
 \hline &&&&&&&  \\[-1.15em]
  Twisted Sector &   $P~;~P+kV_i$& $Q_1$
    &   $Q_2$& $Q_3$ &$Q_X$&$Q_R$ &Mult. of singlets (L or R)  \\
 \hline &&&& \\[-1.4em]
$A(T_9^0)$& $\left(+^9;+,-;-^5\right)$  & $-1$& $-1$ & $+1$& $+5$&$0$ & 1(L) \\ 
 &&&&&&&  \\[-1.25em]
$B(T_9^+)$& $\left(-^9;-,-;+^5\right)$  &  $+1$& $+1$ & $+1$& $-5$&$-3$ & 1(R)  \\ 
 &&&&&&&  \\[-1.25em] 
$C(T_9^-)$& $\left(-^9;-,+;+^5\right)$  &  $+1$& $+1$ & $-1$& $-5$&$0$ &  1(R) \\ 
 &&&&&&&  \\[-1.25em]
\hline &&&& \\[-1.15em]
$D(T_4^+)$& $\left(-^9;+,\pm;+^5\right)$   & $+1$& $-1$ & $\mp 1$& $-5$&$(+2, -1)$ &  3(L)+3(L), 3(R)+3(R) \\ [0.1em]
 && &&&&& \\[-1.25em]
$E(T_4^-)$& $\left(-^9;-,\frac{-3}{2}{\rm ~or~}\frac{-5}{2};-^5\right)$  &$+1$& $-1$ & $+3/+5$& $+5$&$(+6,+3)$ &  7(R)+7(R)  \\ [0.1em]
\hline && &&&&& \\[-1.25em]
 $F(T_1^-)$& $\left(0^9;-1,-1;0^5\right)$  & $0$& $+2$ & $+2$& $0$&$-5$ & 3(L),~neutral\\ [0.15em]
 \hline
$G(T_2^0)$& $\left(0^9;+1,+1;0^5\right)$  &$0$& $-2$ & $-2$& $0$&$+5$ &   3(R),~neutral \\ [0.15em]
   &&&&&&&  \\[-1.25em]
$H(T_2^-)$& $\left(0^9;+1,+1;0^5\right)$  &$0$& $-2$ & $-2$& $0$&$+5$ &   3(R),~neutral \\ [0.15em]
  \hline  &&&&&&&  \\[-1.25em]
  $I\,(T_7^0)$ & $\left(+^9;\frac{-3}{2},-;-^5\right)$  &$-1$& $+3$ & $+ 1$& $+5$&$-4$ &  3(L) \\ [0.25em]
$J(T_7^-)$& $ \left(-^9;+,-;+^5\right)$  &$+1$& $-1$ & $+ 1$& $-5$&$-1$ &  3(L) \\ [0.25em]
\hline
$K_1(T_6^0)$& $\left( +^9;\frac{-3}{2},\frac{-5}{2};+^5\right)$ & $-1$& $+3$ & $+5$& $-5$&$-20$ &  $3\times 7(L),~3 \times 7(R)  $ \\[0.25em] 
$K_2(T_6^0)$& $\left( +^9;\frac{-3}{2},\frac{-7}{2};+^5\right)$ & $-1$& $+3$ & $+7$& $-5$&$-23$ &  $3\times 7(L),~3 \times 7(R)  $ \\[0.25em] 
$K_3(T_6^0)$& $\left( +^9;\frac{-3}{2},\frac{-5}{2};-^5\right)$ &  $-1$& $+3$ & $+5$& $+5$&$-10$ &  $3\times 7(L),~3 \times 7(R)  $ \\[0.25em] 
$K_4(T_6^0)$& $\left( +^9;\frac{-3}{2},\frac{-7}{2};-^5\right)$ & $-1$& $+3$ & $+7$& $+5$&$-13$ &  $3\times 7(L),~3 \times 7(R)  $ \\[0.25em]  
\hline\hline
$T_6 (\ten)$& $(+^9; \frac{-3}{2},  \frac{-5}{2} {\rm ~or~} \frac{-7}{2};\underline{++\,---})$  &  $-1$& $+3$ & $+5/+7$& $+1$&$(-14,-17)$ &  $36(L),~36(R)  $ \\ [0.35em]
$T_6 (\tenb)$& $(+^9; \frac{-3}{2}, \frac{-5}{2} {\rm ~or~} \frac{-7}{2};\underline{+++\,--}) $  &  $-1$& $+3$ & $+5/+7$& $-1$&$(-14,-17)$ &  $36(L),~36(R)  $  \\ [0.35em]
\hline
\end{tabular}
\end{center}
\caption{U(1) charges of $\one$'s and Higgs $\ten$ and $\tenb$. Here,  $Q_R $ is calculated from Eq. (\ref{eq:Rcharges}). }\label{tab:SingR}
\end{table}

\bigskip
\noindent
In Table \ref{tab:Quints}, we collect all $\five$'s, $\fiveb$'s, $\ten$'s, and $\tenb$'s. A set of \anti\,GUT representation, which is of course anomaly-free chiral set, is $\tenb_{-1}\oplus \five_{+3}\oplus\bar\one_{-5}$. Indeed, three $\bar\one_{-5}$'s ($A,B,$ and $C$)  in Table  \ref{tab:SingR} belong to the \anti~families. The remaining $Q_X=\pm 5$ singlets form vector-like representations under \anti. The neutral singlets, $F,G,$ and $H$ are the heavy neutrinos. 

 \begin{table}[!ht]
\begin{center}
\begin{tabular}{|c|c|ccc|c|c|c|}
 \hline &&&&&& & \\[-1.15em]
  States &   $P~;~P+kV_i$ &$Q_1$ &$Q_2$& $Q_3$& $Q_X$  &\,Multiplicities of $\five,\fiveb,\ten$ and $\tenb$& ~$Q_{R}$~ \\
 \hline &&&&  &&& \\[-1.4em]
$\rho\,(\five_{R}(T_1^+))$& $(0^9;+1,0;\underline{-1\,0^4 })$ & $0$&  $-2$& $0$ & $+2$ &   3(R) &   $+4$ \\[0.2em] 
 &&&& && & \\[-1.25em]
 &&&& &&& \\[-1.25em]
$\sigma\,(\fiveb_{R}(T_1^+))$& $(0^9;0,+1;\underline{+1\,0^4 })$ & $0$&  $0$& $-2$ & $-2$ &   3(L) &  $+1$  \\[0.2em]
 &&&& &&  & \\[-1.25em]
 \hline &&&& &&& \\[-1.25em]
$\xi\,(\fiveb_{R}(T_2^-))$& $(-^9;+,+;\underline{-++++}) $ & $+1$&  $-1$& $-1$ & $-3$ &   3(R) &   $+4$  \\[0.2em] 
  &&&& && & \\[-1.25em] 
 &&&& && & \\[-1.25em]
$\eta\,(\five_{R}(T_2^-))$& $(+^9;-,-;\underline{+----}) $ & $-1$&  $+1$& $+1$ & $+3$ &   3(L) &   $-4$ \\[0.2em] 
 &&&& && & \\[-1.25em]
 \hline &&&& &&& \\[-1.25em]
$\alpha\,(\five_{R}(T_1^-))$& $(0^9;-1,0;\underline{+1\,0^4 })$ & $0$&  $+2$& $0$ & $-2$ &   3(R),~Higgs $\five_{hR}$ containing $H_u$&   $-4$\\[0.2em] 
 &&&& && & \\[-1.25em]
 &&&& &&& \\[-1.25em]
$\beta\,(\fiveb_{R}(T_1^-))$& $(0^9;+1,0;\underline{-1\,0^4 }) $ & $0$&  $-2$& $0$ & $+2$ &   3(L),~Higgs $\fiveb_{hR}$ containing  $H_d$&   $+4$\\[0.2em] 
 &&&& && & \\[-1.25em]
 \hline&&&& &&& \\[-1.25em]
$\overline{F}\,(\five_{R}(T_7^-))$& $\left(+^9;+\,+;\underline{+\,-\,-\,-\,-} \right)$ & $-1$&  $-1$& $-1$ & $+3$ &   3(L)& $+1$ \\ 
   &&&& &&&  \\[-1.25em]
$T\,(\tenb_{R}(T_7^-)$)& $\left(-^9;-,-;\underline{+\,+\,+\,-\,-} \right) $ & $+1$&  $+1$& $+1$ & $-1$ &   3(R)&  $+1$\\[0.2em] 
 &&&& &&&  \\[-1.25em]
\hline &&&& &&& \\[-1.25em]
${H} (\ten_R(T_6 ))$ & $\left(-^9;-,- ;\underline{++---}\right) $ &  $+1$& $+1$  & $+1$  & $+1$ & $36(L),~36(R),~VEV= 0$ &  $+3$  \\ [0.35em]
$\overline{H}   (\tenb_R(T_6 ))$& $\left(-^9;-,+ ;\underline{+++--}\right)$&  $+1$ &  $+1$& $-1$   & $-1$& $36(L),~36(R),~ VEV\ne 0$  &  $+4$  \\ [0.15em]
\hline
\end{tabular}
\end{center}
\caption{U(1) charges of $\five,\fiveb,\ten$ and $\tenb$, where $Q_R $ is given in Eq. (\ref{eq:Rcharges}). The L fields are changed to R fields in the first column by taking charge conjugated quantum numbers presented in the text.
}\label{tab:Quints} 
\end{table}

\bigskip
\noindent
From Table \ref{tab:Quints}, the vector-like pairs  $\{\rho,\sigma\}$ and $\{\xi,\eta\}$ are removed. The remaining ones constitute the \anti~spectra $T, \overline{F}, A, B, C$ and the MSSM Higgs fields  $\alpha$ and $\beta$, containing $H_u$ and $H_d$.  The vector-like pairs $H$ and $\overline{H}$ are needed to break the \anti~down to the SM.

\section{Assignment of R-parity}\label{sec:Rparity}
 
\bigskip
\noindent
For a $\Z_{4R}$ symmetry,  we may assign the following U(1)$_{R}$ charges to the MSSM plus $\sigma$ fields,
\begin{eqnarray}
&&q(T(\tenb_{-1,R})), ~\ell(\eta(\five_{+3,R})),~\tau^+(B(\one_{-5,R})),  ~H_u(  \alpha (\five_{-2,R})),  ~H_d(\beta (\fiveb_{+2,R})),~~\sigma \nonumber\\
 Q_{R}:&&\qquad +1,\qquad\qquad +1 ,\qquad\qquad \qquad +1,\qquad \qquad -4,\qquad\qquad\quad +4,\qquad\quad+1   \label{eq:Rcharges}
\end{eqnarray}   
With the four gauged U(1) charges, let us choose $Q_R$ as
\begin{eqnarray}
 Q_R =\frac{9}{2}Q_1-Q_2- \frac32Q_3+Q_X\label{eq:Rcharges}
 \end{eqnarray}
 which is tabulated in  Table \ref{tab:Quints} and also in Table  \ref{tab:SingR}. To break  U(1)$_{R}$ down to $\Z_{4R}$, we need  the GUT breaking VEVs,  $\langle\overline{H}   (\tenb_R(T_6 ))\rangle \ne 0$, but we demand $\langle H(\ten_R(T_6 ))\rangle$ should not develop a VEV such that  $\Z_{4R}$ is not broken further by $\langle H(\ten_R(T_6 ))\rangle$.    With  $\Z_{4R}$, the Yukawa couplings are required to carry the $Q_{R}$ charge 2 modulo 4. With (\ref{eq:Rcharges}), the Higgs quintets carry $Q_{R}=0$ modulo 4 and the Yukawa couplings $T\overline{F}\,\alpha, TT\beta$ and  $\overline{F} B\beta$ are obtained. So,  $B$ is $\tau^+$.  
    
  \subsection{One pair of Higgs quintets}
  
\bigskip 
\noindent     
The Higgs quintets are $\alpha$ and $ \beta$. Twisted sector $T_1^-$ of   $\alpha$ and $ \beta$ gives multiplicity 3. Thus, three objects at these fixed points must have permutation symmetry $S_3$, the largst discrete symmetry of three objects. From three permuting elements of $S_3$, $\{x_1,x_2,x_3\}$,  the  $S_3$ singlet is formed as
\begin{eqnarray}
\left( x_1+  x_2+  x_3\right)/\sqrt3, \label{eq:sofx}
\end{eqnarray}
and the $S_3$ doublet becomes
\begin{eqnarray}
\begin{pmatrix}
\left( x_1+\omega x_2+ \omega^2 x_3\right)/\sqrt3\\[0.5em]
\left( x_1+\omega^2 x_2+ \omega  x_3\right)/\sqrt3
\end{pmatrix}\label{eq:dofx}
\end{eqnarray}
where $\omega$ is the complex cube root of unity.  The tensor product  of two  $S_3$ doublets, $(x_1,x_2)$ and $(y_1,y_2)$, is \cite{Tanimoto10},
\begin{eqnarray}
\two\otimes \two\rightarrow \one\oplus\one'\oplus \two\label{eq:Tprod2}
\end{eqnarray}
where
\begin{eqnarray}
\one= x_1y_1+x_2y_2,~ {\one}'= x_1y_2-x_2y_1,~ \two=
\begin{pmatrix}x_2y_2-x_1y_1\\ x_1y_2+x_2y_1 \end{pmatrix}.
\end{eqnarray}
The Higgsino mass terms, constructed with the  $\alpha$  and  $\beta$ must be an $S_3$ singlet $\one$, but not $\one'$. With this condition, we try to estimate the unmatched pairs of $\five_{hR}$ and $\fiveb_{hR}$, \ie the number of massless Higgsino pairs. Note that SU(9)$'$ confines at  a high energy scale, and we consider the condensation $\tsixb'\cdot\nine'\cdot\nine'$. The $Q_R$ charges of the hidden sector fields given by Eq. (\ref{eq:Rcharges}) are
\begin{eqnarray}
&&\tsixb'(T_9^0), \quad\nine'(T_4^0),\quad\nine'(T_1^-)  \nonumber\\
 Q_{R}:&& ~ +18,\qquad\,  +19 ,\qquad~  +7 \label{eq:Rhidden}
\end{eqnarray}   
The condensation we consider is the scalar component of $\tsixb'$ and the fermion components of $\nine'$. Then, the $Q_R$ charge of
$C\propto \tsixb'_{\vartheta^0}\cdot \nine'_{\vartheta^1}\cdot\nine'_{\vartheta^1}$ are either 54 (both $\nine'$ from $T_4^0$), 30 (both $\nine'$ from $T_1^-$), or 42 (one $\nine'$ from $T_4^0$ and the other from $T_1^-$), \ie $Q_R=2$ modulo 4. Thus, the coupling $\alpha_1\beta_1C$ is allowed and $\alpha_1$ and $\beta_1$ are removed  are removed at the scale where SU(9)$'$ confines, $\langle C\rangle$. Note, however, that $\Z_{4R}$ is broken to $\Z_{2R}$  at the scale $\langle C\rangle$.
  
\bigskip 
\noindent     
Next, let us consider doublets. Two doublets, one each from the $\alpha$ and $\beta$ sets, \ie $\alpha_2$ and $\beta_2$ (the form given in Eq. (\ref{eq:dofx})), combine to form $\one$: $\beta_2\xi_2$,    as noted in  Eq. (\ref{eq:Tprod2}). As above, the coupling $\alpha_2\beta_2C$ is allowed and one of  $\alpha_2$ and one of $\beta_2$ are removed  at the scale where SU(9)$'$ confines, $\langle C\rangle$, and there remains only one pair of  $\five_{hR}$ and $\fiveb_{hR}$. This is the MSSM  vector-like pair of $\five'_{hR}$ and $\fiveb'_{hR}$.  Summarizing, 
 \begin{eqnarray}
\five_{hR}\oplus \fiveb_{hR}:~\textrm{one  MSSM pair}, ~\textrm{two pairs at the scale} ~\langle C\rangle.
\end{eqnarray}
From two permuting elements $x_2$ and $x_3$, the $S_2$ singlet is $(x_2+x_3)/\sqrt2$, and the orthogonal component to that is  $(x_2-x_3)/\sqrt2$. Thus, the MSSM doublets are
\begin{equation}
\frac{1}{\sqrt2}\left(\five_{hR}(T_1^-)_2-\five_{hR}(T_1^-)_3\right), ~\frac{1}{\sqrt2}\left(\fiveb_{hR}(T_1^-)_2-\fiveb_{hR}(T_1^-)_3\right).
\end{equation}

\subsection{The third family members}
 
\bigskip 
\noindent    
It is natural to choose the singlet combination $(x_1+x_2+x_3)/\sqrt3$ as the third family member, and the doublet as the light family members. The quark and lepton mass matrix can be constructred based on these bases.

\section{Conclusion}\label{Conclusion}

\noindent
We derived a $\Z_{4R}$ parity in a family unification model with a GUT  \anti~possessing three families and one pair of Higgs quintets. Spontaneous symmetry breaking of anti-SU(5) GUT is achieved by   $\langle \ten_{+1}\rangle \oplus\langle \tenb_{-1}\rangle $. The $\Z_{4R}$ parity together with the permutation symmetry $S_3$ are useful to select the third family members and one MSSM pair of the Higgs quintets.

\acknowledgments{\noindent I thank the APCTP for the hospitality extented to me during the visit. This work is supported in part  by the National Research Foundation (NRF) grants  NRF-2018R1A2A3074631.}
      
\begin{appendix}

\section{Fixed points in $\Z_{12-I}$}\label{app:Fixed}
 \bigskip
 \noindent
In the $k$-th twisted sector of $\Z_{N}$ orbifold, multiplicities  ${\cal P}_k$  is
\begin{eqnarray}
{\cal P}_k   =\frac{1}{N}\sum_{l=0}^{N}
\tilde\chi(k,l)e^{i\,2\pi l\Theta_k},\label{eq:Multiplicity}
\end{eqnarray}
where $\tilde{\chi}(k,l)$ in the $\Z_{12-I}$ orbifold are listed in  
\begin{table}[!ht]
\begin{center}
\begin{tabular}{|c|cccccccccccc|}
 \hline 
 &   & & & & &   $l$& $=$& & & &  &\\[-0.3em]
 $k$ &0&1&2 &3&4&5&6&7&8&9&10& 11\\ 
 \hline
 1&  3&3&3&3 &3&3&3 &3&3&3 &3&3\\[-0.2em] 
 2 &  3&3&3&3 &3&3&3 &3&3&3 &3&3\\[-0.2em] 
3 & 4& 1& 1& 4& 1& 1&  4& 1& 1& 4& 1& 1 \\ [-0.2em]
4& 9&1&1&1 &9&1&1 &1&9&1 &1&1 \\[-0.2em] 
   5 &  3&3&3&3 &3&3&3 &3&3&3 &3&3\\[-0.2em] 
6 & 16&1&1&4 &1&1&16 &1&1&4 &1&1\\
\hline
\end{tabular}
 \caption{ $\tilde{\chi}(k,l)$ in the $\Z_{12-I}$ orbifold \cite{LNP954}. }\label{tab:Chis12I}
\end{center} 
\end{table}
Table  \ref{tab:Chis12I}. In the calculation of multiplicity the information on $\Theta$ is the key. The right-hand side (RHS) mover is denoted by the spinor  
\begin{eqnarray}
s=(s_0;\tilde{s})=(\ominus ~{\rm or}~\oplus\,;\pm,\pm,\pm),\label{eq:sDef}
\end{eqnarray}
where $s_0$ corresponds to L- or R- movers. In this paper, $+$ or $-$ in the spinor form as in Eqs. (\ref{eq:VtwoShift}) and (\ref{eq:sDef}), denote $+\frac12$ and  $-\frac12$, respectively. The phase $\Theta_k$ in the $k^{\rm th}$ twisted sector  in  is   
\begin{eqnarray}
&\Theta_k  = \sum_j (N^j_L-N^j_R)\hat{\phi}^j -\frac{k}{2}(V_a^2-\phi_s^2)+(P+kV_a)\cdot V_a-(\tilde s +k\phi_s)\cdot\phi_s +{\rm integer}, \nonumber\\
&=-\tilde{s}\cdot\phi_s+\Delta_k,\label{eq:Phase}
\end{eqnarray}
where  $\Delta_k$ is
\begin{eqnarray}
&&\Delta_k = (P+kV_a)\cdot V_a-\frac{k}{2}(V_a^2-\phi_s^2)+\sum_j (N^j_L-N^j_R)\hat{\phi}^j\\
&&\qquad\equiv\Delta_k^0+\Delta_k^N.\label{eq:Deltak}
\end{eqnarray}
$V_a$ is the shift vector $V$ distinguished by Wilson lines $a,\,V_{0,+,-}$, and
\begin{eqnarray}
&&\Delta_k^0 =P\cdot V_a+\frac{k}{2}(-V_a^2+\phi_s^2) , \\
&&\Delta_k^N =\sum_j (N^j_L-N^j_R)\hat{\phi}^j,\label{eq:PhaseMore}
\end{eqnarray}
and $\hat{\phi}^j$ is in the range $0<\hat{\phi}^j\le 1$ mod integer.  The oscillator contributions in $\Delta_k^N$ is  for $N_{L,R}\ge 0$. We satisfy the massless spectrum condition for gauge sectors by
\begin{eqnarray}
(P+kV_a)^2+2\sum_j N_L^j \hat{\phi}^j = 2\tilde{c}_k ,\label{eq:vacuumE}
\end{eqnarray}
where $2\tilde{c}_k$ of $\Z_{12-I}$ orbifold is
\begin{eqnarray}
\begin{array}{ll}
2\tilde{c}_k:  &
 ~ \frac{210}{144}(k=1),~ \frac{216}{144}(k=2),~ \frac{234}{144}(k=3),~ \frac{192}{144}(k=4),~ \frac{210}{144}(k=5),~ \frac{216}{144}(k=6), 
 \end{array} \label{eq:Twist121}
\end{eqnarray}
which will lead to ${\cal N}=1$ supersymmetry in 4D.  
 
\bigskip
 \noindent
In the twisted sectors with with $3a_3=0$ mod. integer, \ie at $ T_{3,6,9}$ and also $U$  \cite{KimKyae07},\footnote{$T_9$ contains the CTP conjugate states of $T_3$.}  
the gauge invariance condition has to be taken into account explicitly,
\begin{eqnarray}
(P+kV_0)\cdot a_3=0.\label{eq:WilsonCond}
 \end{eqnarray}
 We distinguish the twisted sector $T_k$ by $0,+$ and $-$. The phase  $\Delta_k^0$ of Eq. (\ref{eq:PhaseMore}) contains an extra $\frac{k}{2}$ factor, but at $T_6$ the factor  $\frac{k}{2}$ is an integer. At $T_{1,2,4,5}$, we distinguish three fixed points just by $V_{0,+,-}$.
 
\section{Useful tables for model building}\label{app:Tables}
\noindent
Here we present multiplicities in Table \ref{tab:Multiplicities} and $H$-momenta in Table \ref{tab:deltak} \cite{LNP954}.

\begin{table}[!ht]
\begin{center} 
 \begin{tabular}{|c |ccccc|}
\hline
  &&&{\rm Multiplicity}&\\[-1.4em]
  &&&&&\\
  i  & ${\cal P}_k (0)$ &~ ${\cal P}_k (\pm\frac{\pi}{3})$& ${\cal P}_k ( \pm\frac{2\pi}{3} )$& ${\cal P}_k (\pm \frac{ \pi}{2} )$ &~~ ${\cal P}_k (\pi)$   \\[0.3em]
 \hline
 1   &3 &0&0&0&0\\
 2  & 3 &0 &0&0&0 \\
 3  &2 &0&1&0&0\\
 4  &3 &0&0&2&2\\
 5  &3&0&0&0&0\\
 6  &{4}  &2 &3 &0 &{2} \\
\hline
  \end{tabular}
\end{center}
\caption{Multiplicities in the $k$-th twisted sectors of   $\Z_{12-I}$.  ${\cal P}_k({\rm angle})$ is calculaed with angle\,$=\frac{2\pi}{12}\cdot l$ in Eq. (\ref{eq:Multiplicity}).}\label{tab:Multiplicities}
\end{table}

 \begin{table} [!ht]
\begin{center}
\begin{tabular}{|c|c|ccc|cc| }
 \hline  
 Orbifold&\,Twisted Sector\,&  ~$k\,\hat{\phi}$~ & ~ $p_{\rm vec}$ & $p_{\rm orb}$& $-k\,p_{\rm vec}^{k\,\rm th}\cdot {\phi}_s$ & ~$\delta_k$~ \\[0.2em] \hline  
&$T_1$   & ~$(\frac{5}{12},\frac{4}{12} ,\frac{1}{12})$ &$(-1,0 ,0) $&$(\frac{-7}{12},\frac{+4}{12} ,\frac{+1}{12}) $ & $ \frac{5}{12}$ & $ \frac{1}{12}$   \\[0.3em]  
&$ T_2$    &~$(\frac{5}{6},\frac{4}{6}, \frac{1}{6})$&$(-1,0 ,0)$&   $(\frac{-1}{6},\frac{+4}{6} ,\frac{+1}{6})$ & $ \frac{10}{12}$& 0 \\[0.3em]   
\,$\Z_{12-I}$\, &~$ T_3$ ~ & ~$ (\frac{5}{4},\frac{4}{4} ,\frac{1}{4})$ &$ (-1,-1 ,-1)$
 & $(\frac{1}{4},\frac{0}{4} ,\frac{-3}{4})$& $ \frac{6}{12}$ &$\frac{3}{12}$ \\[0.3em]  
&$ T_4$   & ~$(\frac{5}{3},\frac{4}{3} ,\frac{1}{3})$&$ (-2,-1,0)$ &$(\frac{-1}{3},\frac{1}{3} ,\frac{1}{3})$& $ \frac{8}{12}$ &0  \\[0.3em]
 &$ T_5$ &~$(\frac{25}{12},\frac{20}{12} ,\frac{5}{12})$ &$(-2,-2,-1)$ &$(\frac{1}{12},\frac{-4}{12} ,\frac{-7}{12})$
  & $\times$   &$\times$ \\[0.3em]
&$ T_6$   &~$(\frac{5}{2},\frac{4}{2} ,\frac{1}{2}) $ &$(-2,-2,0) $ &$ (\frac{1}{2},\frac{0}{2} ,\frac{1}{2}) $ &0&0 \\[0.3em] 
 &$ T_7$   &~$(\frac{35}{12},\frac{28}{12} ,\frac{7}{12}) $ &$(-3,-1,0) $ &$ (\frac{-1}{12},\frac{4}{12} ,\frac{7}{12}) $   &$\frac{1}{12}$  &$\frac{1}{12}$ 
  \\[0.3em] 
&$ T_8$   &~$(\frac{40}{12},\frac{32}{12} ,\frac{8}{12}) $ &$(-3,-3,-1) $ &$ ( \frac{1}{3},\frac{-1}{3} ,\frac{-1}{3}) $ &$\frac{8}{12}$     &0   \\[0.3em] 
&$ T_9$   &~$(\frac{45}{12},\frac{36}{12} ,\frac{9}{12}) $ &$(-4,-3,-1) $ &$ (\frac{-3}{12},\frac{0}{12} ,\frac{-3}{12}) $   &$\frac{-3}{12}$    &$\frac{3}{12}$   \\[0.3em] 
 \hline
\end{tabular}
\end{center}
\caption{$H$ momenta, $p_{\rm orb}$, in the twisted sectors of $\Z_{12-I}$, Table 10.1 of \cite{LNP954}.   Requiring $(p_{\rm vec}+k\hat{\phi})^2=$(2nd line in Eq. (\ref{eq:Twist121})),  we have $p_{\rm vec}$ in the 4th column. In the last column, $\delta_k$ is shown, from which we have  the energy contribution from right movers $2\delta_k\ge 0$. } \label{tab:deltak}
\end{table}

\section{Multiplicities for each $\hat{s}$}\label{app:Phihat}
\noindent
The multiplicities are calculated for each $\hat{s}$.   In Table \ref{tab:ThSix}, we present one example for a complete list of the RHS spin vectors $\hat{s}$, deriving $\tsix$ of SU(9)$'$ in  $T_9^0$.

 \begin{table}[!ht]
\begin{center}
\begin{tabular}{|cc|c|cc|ccc|cc|}
 \hline &&&&&&&&& \\[-1.15em]
  Chirality  &   $\tilde s$& $-\tilde{s}\cdot\phi_s$& $-k\,p_{\rm vec}^{ k\,\rm th}\cdot \phi_s$ &$k\,P_{9}\cdot V_0$& $(k/2)\phi_s^2$,&$  -(k/2) V_0^2 $,& ~~$\Delta_{9}^N,~-\delta_9^N
  $   &~~$\Theta_{9},$& Mult. of SU(9)  \\[0.15em]
 \hline &&&&&&& && \\[-1.15em]
$\ominus=L$& $(---)$  &  $\frac{+5}{12}$ & $~~\frac{-3}{12},$& $\frac{+6}{12}$ &$\frac{189}{144}  $  &$\frac{+243}{144}$ & $ \frac{0}{12},~~\frac{- 6}{12}$&$\frac{+2}{12} +\frac{-2}{12}  $& 2  \\ [0.1em]
 \hline &&&&&&&&& \\[-1.25em]
$\ominus=L$& $(-++)$  &  $\frac{0}{12}$ & $~~\frac{-3}{12},$& $\frac{+6}{12}$ &$\frac{189}{144}  $  &$\frac{+243}{144}$ & $ \frac{0}{12},~~\frac{- 6}{12}$&$\frac{-5}{12}    $& 0 \\ [0.1em]
 \hline &&&&&&&&& \\[-1.25em]
$\ominus=L$& $(+-+)$  &  $\frac{-1}{12}$ & $~~\frac{-3}{12},$& $\frac{+6}{12}$ &$\frac{189}{144}  $  &$\frac{+243}{144}$ & $ \frac{0}{12},~~\frac{- 6}{12}$&$\frac{-6}{12} $& 0  \\ [0.1em]
 \hline &&&&&&&&& \\[-1.25em]
 $\ominus=L$& $(++-)$  &  $\frac{-4}{12}$ & $~~\frac{-3}{12},$& $\frac{+6}{12}$ &$\frac{189}{144}  $  &$\frac{+243}{144}$ & $ \frac{0}{12},~~\frac{- 6}{12}$&$\frac{-9}{12}  $& 0  \\ [0.1em]
 \hline &&&&&&&&& \\[-1.25em]
 $\oplus=R$& $(+++)$  &  $\frac{-5}{12}$ & $~~\frac{-3}{12}$&$\frac{+6}{12}$  &$\frac{189}{144}$ &$\frac{+243}{144}$&$ \frac{0}{12},~~\frac{-6}{12}$& $\frac{-10}{12}$  &0   \\[0.15em] 
\hline
$\oplus=L$& $(+--)$  &  $\frac{0}{12}$ & $~~\frac{-3}{12},$& $\frac{+6}{12}$ &$\frac{189}{144}  $  &$\frac{+243}{144}$ & $ \frac{0}{12},~~\frac{- 6}{12}$&$\frac{-5}{12}  $&0  \\ [0.1em]
 \hline &&&&&&&&& \\[-1.25em]
$\oplus=L$& $(-+-)$  &  $\frac{+1}{12}$ & $~~\frac{-3}{12},$& $\frac{+6}{12}$ &$\frac{189}{144}  $  &$\frac{+243}{144}$ & $ \frac{0}{12},~~\frac{- 6}{12}$&$\frac{-4}{12} $& 1  \\ [0.1em]
 \hline &&&&&&&&& \\[-1.25em]
$\oplus=L$& $(--+)$  &  $\frac{+4}{12}$ & $~~\frac{-3}{12},$& $\frac{+6}{12}$ &$\frac{189}{144}  $  &$\frac{+243}{144}$ & $ \frac{0}{12},~~\frac{- 6}{12}$&$\frac{-1}{12}$& 0  \\ [0.1em]
 \hline  
\end{tabular}
\end{center}
\caption{The multiplicities for each $\hat{s}$ calculated with the row $i=3$ in Table \ref{tab:Multiplicities}.  }\label{tab:ThSix}
\end{table}

\end{appendix}


\newpage

\end{document}